\documentclass{article}
\usepackage{spconf}
\usepackage{blindtext, graphicx}
\usepackage{bm}
\usepackage{booktabs}
\usepackage{subfig}
\usepackage[font=footnotesize]{caption}
\usepackage{hyperref}

\usepackage{amsmath}
\usepackage{amssymb}
\usepackage{amsthm}
\usepackage{latexsym}
\usepackage{verbatim}
\usepackage{graphicx}
\usepackage{mdframed}

\definecolor{shade}{rgb}{1,1,1}

{\begin{list}               
    {$\bullet$ \hfill}{
        \setlength{\leftmargin}{\parindent}
        \setlength{\parsep}{0.04\baselineskip}
        \setlength{\itemsep}{0.5\parsep}
        \setlength{\labelwidth}{\leftmargin}
        \setlength{\labelsep}{0em}}
    }
{\end{list}}

\providecommand{\cref}[1]{Chapter~\ref{chap:#1}}

\providecommand{\R}{\ensuremath{\mathbb{R}}}
\providecommand{\C}{\ensuremath{\mathbb{C}}}

\providecommand{\abs}[1]{\left|#1\right|}

\providecommand{\set}[1]{\left\lbrace#1\right\rbrace}

\providecommand{\diag}{\mathop{\mathrm{diag}}}

\providecommand{\trace}{\mathop{\mathrm{trace}}}

\renewcommand{\vec}[1]{\ensuremath{\mathbf{#1}}}
\providecommand{\mat}[1]{\ensuremath{\mathbf{#1}}}

\providecommand{\wh}[1]{\ensuremath{\widehat{#1}}}

 \providecommand{\mD}{\mat{D}}
 \providecommand{\mH}{\mat{H}}

 \providecommand{\mP}{\mat{P}}

\providecommand{\mV}{\mat{V}} 

\providecommand{\mSigma}{\mat{\Sigma}}

\providecommand{\mY}{\mat{Y}}
\providecommand{\mZ}{\mat{Z}}

\providecommand{\conv}{\ast}

\renewcommand{\vec}[1]{\mathsf{#1}}

\newcommand{\vY}{\vec{Y}}
\newcommand{\vX}{\vec{X}}
\newcommand{\vB}{\vec{B}}
\newcommand{\vQ}{\vec{Q}}

\begin{document}
\ninept

\title{Separake: Source separation with a little help from echoes}

\name{Robin Scheibler$^\dagger$, Diego Di Carlo$^\ddagger$, Antoine Deleforge$^\ddagger$, and Ivan Dokmani\'{c}$^\sharp$\thanks{
The research presented in this paper is reproducible. Code and data are available at \protect\url{dokmanic.ece.illinois.edu/separake.zip}. Ivan Dokmani\'c was supported by a Google Faculty Research Award.}}
\address{$^\dagger$Tokyo Metropolitan University, Tokyo, Japan\\
         $^\ddagger$Inria Rennes - Bretagne Atlantique, France\\
         $^\sharp$Coordinated Science Lab, ECE, University of Illinois at Urbana-Champaign}

%

\maketitle

\begin{abstract}
It is commonly believed that multipath hurts various audio processing algorithms. At odds with this belief, we show that multipath in fact helps sound source separation, even with very simple propagation models. Unlike most existing methods, we neither ignore the room impulse responses, nor we attempt to estimate them fully. We rather assume that we know the positions of a few virtual microphones generated by echoes and we show how this gives us enough spatial diversity to get a performance boost over the anechoic case. We show improvements for two standard algorithms---one that uses only magnitudes of the transfer functions, and one that also uses the phases. Concretely, we show that multichannel non-negative matrix factorization aided with a small number of echoes beats the vanilla variant of the same algorithm, and that with magnitude information only, echoes enable separation where it was previously impossible.
\end{abstract}


\section{Introduction}

Source separation algorithms can be grouped according to how they deal with sound propagation: those that ignore it \cite{le2015deep}, those that assume a single anechoic path \cite{rickard2007duet}, those that model the room transfer functions (TFs) entirely \cite{ozerov2010multichannel,nugraha2016multichannel}, and those that attempt to separately estimate the contribution of the early echoes and the contribution of the late tail \cite{leglaive2015multichannel}. In this paper we propose yet another route: we assume knowing the locations of a few walls relative to the microphone array, which enables us to exploit the associated \textit{virtual microphones}. This assumption is easy to satisfy in living rooms and conference rooms, but the corresponding model incurs a significant mismatch with respect to the complete reverberation. We show that it nonetheless gives sizable performance boosts while being simple to estimate.

\begin{figure}
    \centering
    \includegraphics[width=.7\linewidth]{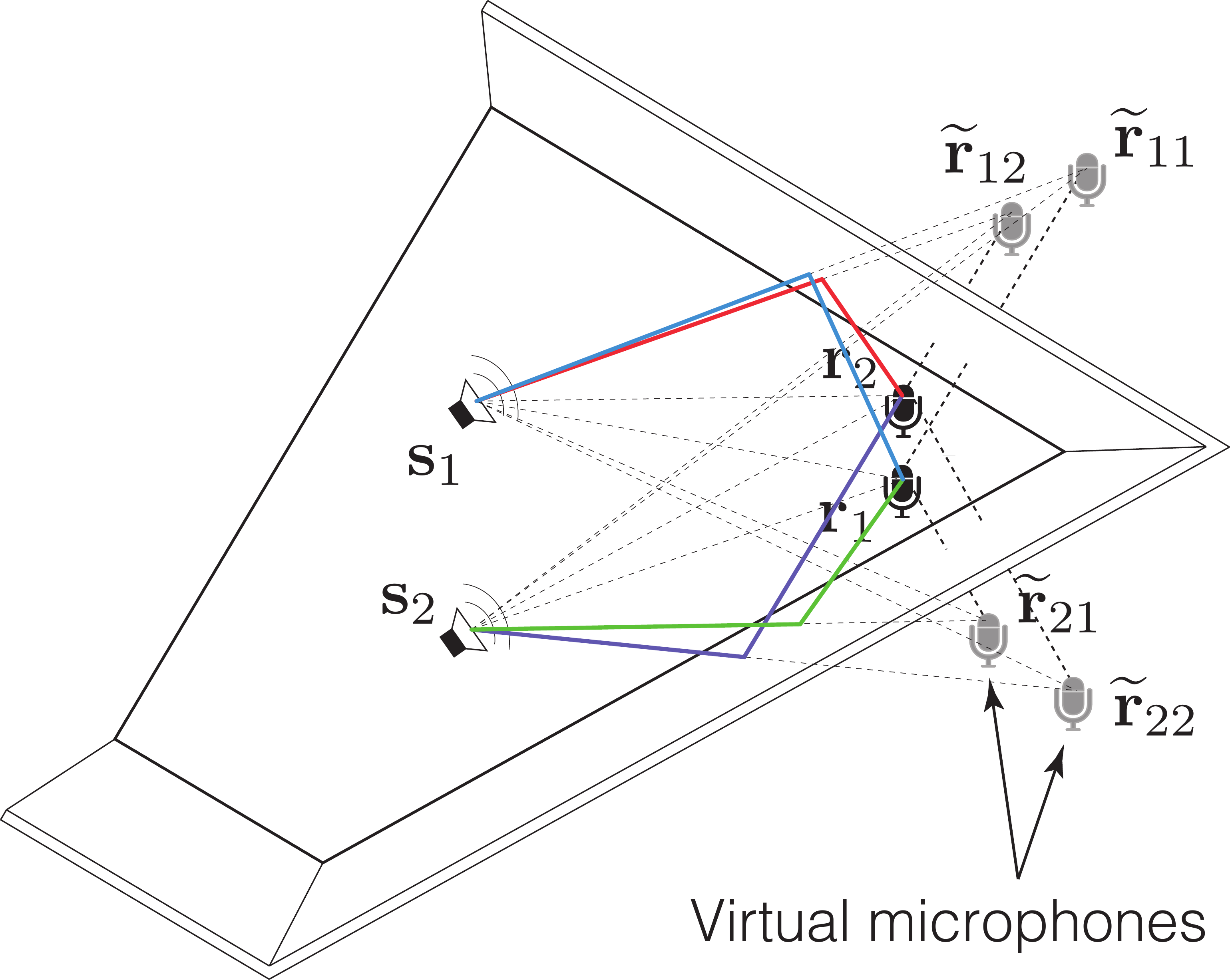}
    \caption{Typical setup with two speakers recorded by two microphones. The illustration shows the virtual microphone model (grey microphones) with direct sound path (dotted lines) and resulting first-order echoes (colored lines)}
    \label{fig:setup}
    \vspace{-0mm}
\end{figure}

A typical setup is illustrated in Figure \ref{fig:setup}. We consider $J$ sources emitting from $J$ distinct directions of arrival (DOAs) $\set{\theta_j}_{j=1}^J$, and an array of $M$ microphones. The array is placed close to a wall or a corner. There are two reasons why this is useful: first, it renders echoes from the nearby walls significantly stronger than all other echoes; second, it keeps the resulting virtual array (real and virtual microphones) compact. The latter justifies the far field assumption which in turn simplifies exposition.

Real and virtual microphones form dipoles with diverse frequency-dependent directivity patterns. Our goal is to design algorithms which benefit from this known spatial diversity.

Echoes have been used previously to enhance various audio processing tasks. It was shown that they improve indoor beamforming \cite{Dokmanic:2015dr, Scheibler:2015ii, RobinThesis}, aid in sound source localization \cite{Ribeiro:2010uj}, and enable low-resource microphone array self-localization \cite{Dokmanic:2016gu}. They, however, seems to be rarely analyzed in the context of source separation with non-negative source models. Since most speech communication takes place indoors, we believe that our findings are relevant for modern applications and voice-based assistants like Amazon Echo or Google Home.

\subsection{Our Goal and Main Findings}

Our emphasis here is different than that in \cite{leglaive2015multichannel}. Rather than fit the echo model, we aim to show that separation in the presence of echoes is in fact better than separation without echoes. We ask the following questions:
\begin{enumerate}
    \item Is speech separation with echoes fundamentally easier than speech separation without echoes? Are there specific settings where this is true or false?
    \item Is it necessary to fully model the reverberation or can we get away with a geometric perspective where we know the locations of a few virtual microphones?
\end{enumerate}
To answer these questions we set up several simple experiments. We take two standard, well-understood multi-channel source separation algorithms which estimate the channel (the TFs), and instead of updating the channel estimate we simply furnish the TFs of real and a few virtual microphones. The first algorithm---non-negative matrix factorization (NMF) via multiplicative updates (MU)---only uses the magnitudes of the transfer functions, while the second one---expectation maximization (EM)---also uses the phases. In this initial investigation we look at the (over)determined case ($J \leq M$); the analysis of the underdetermined case is postponed to a forthcoming journal version of this paper. Our findings can be summarized as follows:
\begin{itemize}
    \item (MU) With magnitudes only, multi-channel anechoic separation is hardly any better than single-channel separation: as the magnitude of the transfer functions is the same at all microphones, channel modeling offers no diversity. The situation is different in rooms where the direction-dependent magnitude of TFs varies significantly from microphone to microphone. \textit{We show that replacing the transfer functions with a few echoes (even just one) gives significant performance gains compared to not modeling the TFs at all, but also that it does better than learning the TFs through multiplicative updates.}
    \item (EM) With both phases and magnitudes, anechoic separation will be near-perfect since it corresponds to a determined linear system. Therefore, any uncertainty from imperfections in channel modeling will make things worse. \textit{Surprisingly, approximating the TFs with one echo matches learning them through EM updates and using more outperforms it.}
\end{itemize}
For a sneak peak at the gains, fast forward to Figure \ref{fig:results}.

\section{Modeling}

Suppose $J$ sources emit inside the room and we have $M$ microphones. Each microphone receives
\[
    y_m(t) = \sum_{j = 1}^J c_{jm}(t),
\]
with $c_{jm}$ being the spatial image of the $j$th source at the $m$th microphone. Spatial images are given as
\[
    c_{mj}(t) = (x_j \conv h_{jm})(t),
\]
where $h_{jm}$ is the room impulse response between the source $j$ and microphone $m$. The room impulse response is a central object in this paper. We model it as
\[
    h_{jm}(t) = \sum_{k = 0}^K \alpha_{jm}^k \delta(t - t_{jm}^k) + e_{jm}(t),
\]
where the sum comprises the line-of-sight propagation and the earliest $K$ echoes we want to account for (at most 6 in this paper), while the error term $e_{jm}(t)$ collects later echoes and the tail of the reverberation. We do not assume $e_{jm}(t)$ to be known. We assume that the sources are in the far field of real and virtual microphones so the times $t_{jm}^k$ depend only on the source DOAs which we assume are known.  Assuming $K$ echoes per source are known, we can form an approximate TF from source $j$ to microphone $m$,
\begin{equation}
    \label{eq:approx_tf}
    \wh{H}_{j,m}(e^{j\omega}) = \sum_{k=0}^K \alpha_{jm}^k e^{-i \omega t_{jm}^k}.
\end{equation}
The far field assumption implies that only the relative arrival times are known so we can arbitrarily fix the delay of the direct path to zero. In addition, we assume all walls to be spectrally flat in the frequency range of interest and set all $\alpha_{jm}^k$ to a same constant.

As usual, we process by frames. In the short-time Fourier transform (STFT) domain the $m$th microphone signal reads
\begin{equation}
    \label{eq:stft_mixing}
    Y_m[f,n] = \sum_{j = 1}^J \wh{H}_{jm}[f] X_{j}[f,n] + B_m[f,n]
\end{equation}
with $f$ and $n$ being the frequency and frame index, $X_{j}[f,n]$ the STFT of the $j$th source signal, and $B_m[f,n]$ a term including noise and model mismatch. It is convenient to group the microphone observations in vector form,
\begin{equation}
    \vY[f,n] = \wh{\mH}[f] \, \vX[f,n] + \vB[f,n].
\end{equation}
where $\vY[f, n] = \big[ \, Y_m[f, n] \, \big]_m$, $\wh{\mH}[f] = \big[ \, \wh{H}_{jm}[f, n] \, \big]_{m,j}$, $\vX[f, n] = \big[ \, X_j[f, n] \, \big]_j$, and $\vB[f, n] = \big[ \, B_m[f, n] \, \big]_m$.
Let the squared magnitude of the spectrogram of the $j$th source be $\mP_j = \big[ \abs{X_{j}[f, n]}^2 \big]_{fn}$. We postulate a non-negative factor model for $\mP_j$: 
\begin{equation}
    \label{eq:nmf_model}
    \mP_j =  \mD_j \mZ_j,
\end{equation}
where $\mD_j$ is the non-negative \textit{dictionary}, and the latent variables $\mZ_j$ are called \textit{activations}. Source separation can then be cast as an inference problem in which we maximize the likelihood of the observed $\vY$ over all possible non-negative factorizations \eqref{eq:nmf_model}. This normally involves learning the channel (frequency-domain mixing matrices). Instead of learning, we fix the channel to the earliest few echoes.

\section{Source Separation by NMF}

To evaluate the usefulness of echoes in source separation, we modify the
multi-channel NMF framework of Ozerov and F\'{e}votte \cite{ozerov2010multichannel} as follows. First, we 
introduce a dictionary learned from available training data. We explore both speaker-specific and universal dictionaries
\cite{Sun:2013co}. Speaker-specific dictionaries can be beneficial when speakers are known in advance. Universal dictionary
is more versatile but gives a weaker regularization prior. Second, rather than learning the TF from the data,
we use the approximate model of \eqref{eq:approx_tf}. In the following we briefly describe the two used algorithms.

\subsection{NMF using Multiplicative Updates (MU-NMF)}\label{sec:mu}

Multiplicative updates for NMF only involve the magnitudes and are simpler than the EM updates. They have been originally proposed by Lee and Seung \cite{Lee:2001ti}. We use the Itakura-Saito divergence \cite{Fevotte:2011af} between the observed multi-channel squared magnitude spectra $\mV_m = [|Y_m[n,f]|^2]_{fn}$ and their non-negative factorizations,
\begin{equation}
    \label{eq:mu_nmf_model}
    \wh{\mV}_m = \sum_{j=1}^{J} \diag(\vQ_{jm}) \mD_j \mZ_j, \quad m=1,\ldots,M
\end{equation}
where $\vQ_{jm} = \big[ \, |\wh{H}_{jm}[f]|^2 \,\big]_f$ is the vector of squared magnitude of the approximate TF between microphone $m$ and source $j$.
We add an $\ell_1$-penalty term to promote sparsity in the activations due to the potentially large size of the universal dictionary~\cite{Sun:2013co}.
The cost function is thus
\begin{equation}
    C_{\mathsf{MU}}(\mZ_j) = \sum_{mfn} d_{\mathsf{IS}}(V_{m}[f,n] | \wh{V}_{m}[f,n]),
    + \gamma \sum_j \| \mZ_j \|_1,
\end{equation}
where $d_{\mathsf{IS}}(v | \hat{v}) = \frac{v}{\hat{v}} - \log \frac{v}{\hat{v}} - 1$. By adapting the original MU rule derivations from Ozerov and F\'{e}votte, we obtain the following regularized MU update rule:
\begin{align}
    \mZ_j \gets \mZ_j \odot \frac{\sum_m (\diag(\vQ_{ij}) \mD_j)^\top \left(\mV_j \odot \wh{\mV}_j^{-2}\right)}{\sum_m(\diag(\vQ_{ij}) \mD_j)^\top \wh{\mV}_j^{-1} + \gamma},
\end{align}
where multiplication $\odot$, power, and division are element-wise.

Importantly, neglecting the reverberation (or working in the anechoic regime) leads to a constant $\vQ_{jm}$ for all $j$ and $m$. A consequence is that the MU-NMF framework breaks down with a universal dictionary. Indeed, \eqref{eq:mu_nmf_model} becomes the same for all $m$, 
$
    \wh{\mV}_m = \sum_{j} \mD \mZ_j = \mD \sum_j \mZ_j,
$
so even with correct atoms chosen, we can assign them to any source without changing
$\mY_m$, and hence the cost. Therefore, anechoic multi-channel separation with a universal dictionary cannot work well. This intuitive reasoning is corroborated by numerical experiments in Section \ref{sec:results}. The problem is overcome by the EM-NMF algorithm which keeps the channel phase and is thus able to exploit the phase diversity across the array. Of course, in line with the message of this paper, it is also overcome by using echoes.

\subsection{NMF using Expectation Maximization (EM-NMF)}

Unlike the MU algorithm that independently maximizes the log-likelihood of TF magnitudes, EM-NMF maximizes the joint log-likelihood over all complex-valued channels~\cite{ozerov2010multichannel}. Hence, it takes into account observed phases.
Each source $j$ is modeled as the sum of components with complex Gaussian priors of the form $c_{k}[f,n]\sim \C\mathcal{N}\Big(0, d_{fk}z_{kn}\Big)$ such that
\begin{equation}
    X_j[f,n] \sim \C\mathcal{N}\left(0, (\mD_j\mZ_j)_{fn}\right),
\end{equation}
and the magnitude spectrum $\mP_j$ of \eqref{eq:nmf_model} can be understood as the variance of source $j$.
Under this model, and assuming uncorrelated noise, the microphone signals also
follow a complex Gaussian distribution with covariance matrix
\begin{equation}
    \mSigma_{\vY}[f,n] = \wh{\mH}[f] \, \mSigma_\vX[f,n] \, \wh{\mH}^H[f] + \mSigma_\vB[f,n],
\end{equation}
and the negative log-likelihood of the observed signal is
\begin{equation}\nonumber 
\resizebox{\linewidth}{!}{
    $C_{\mathsf{EM}}(\mZ_j) = \sum\limits_{fn} \trace\left(\vY[f,n]\vY[f,n]^H\mSigma_{\vY}^{-1}[f,n]\right) \\
    + \log\det\mSigma_{\vY}[f,n].$
    }
\end{equation}
This quantity can be efficiently minimized using the EM algorithm proposed in~\cite{ozerov2010multichannel}. We modify the original algorithm by fixing the source dictionaries $\mD_j$ and the early-echo channel model $\wh{\mH}[f]$ throughout the iterations.

\section{Numerical Experiments}

We test our hypotheses through computer simulations. In the following, we describe the simulation setup, dictionary learning protocols, and we discuss the results.

\subsection{Setup}

\begin{figure}
    \centering
    \includegraphics{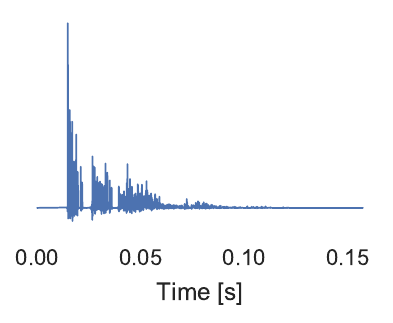}
    \includegraphics{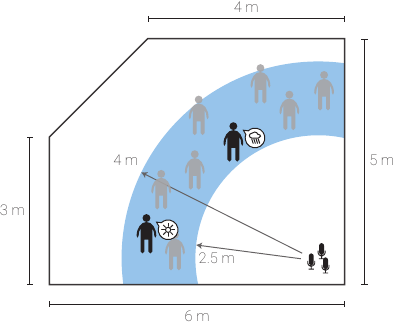}
    \caption{On the left, a typical simulated RIR. On the right, the simulated scenario.}
    \label{fig:rir_room}
\end{figure}

An array of three microphones arranged on the corners of an equilateral triangle with edge length 0.3~m is placed in the corner of a 3D room with 7 walls. We select 40 sources at random locations at a distance ranging from 2.5~m to 4~m from the microphone array. Pairs of sources are chosen so that they are at least 1~m apart. The floor plan and the locations of microphones are depicted in Figure~\ref{fig:rir_room}. The scenario is repeated for every two active sources out of the 780 possible pairs.

The sound propagation between sources and microphones is simulated using the
image source model implemented in \textit{pyroomacoustics} Python package~\cite{scheibler2017pyroomacoustics}. The wall absorption factor is set to 0.4, leading to a T60 of approximately 100~ms. An example RIR is shown in Figure~\ref{fig:rir_room}. The sampling frequency is set to 16~kHz, STFT frame size to 2048 samples with 50\% overlap between frames, and we use a cosine window for analysis and synthesis. Partial TFs are then built from the $K$ nearest image microphones. The global delay is discarded. 


With this setup, we perform three different experiments. In the first one, we evaluate MU-NMF with a universal dictionary. In the other two, we evaluate the performance of MU-NMF and EM-NMF with Speaker-specific dictionaries. We vary $K$ from 1 to 6 and use three baseline scenarios:
\begin{enumerate}
\item \textit{learn}: The TFs are learned from the data along the activations as originally proposed~\cite{ozerov2010multichannel}.
\item \textit{anechoic}: Anechoic conditions, without echoes nor model mismatch.
\item \textit{no echoes}: Reverberation is present but ignored (i.e. $K=0$).
\end{enumerate}
With the universal dictionary, the large number of latent variables warrants the introduction of sparsity-inducing regularization. The value of the regularization parameter $\gamma$ was chosen by a grid search on a holdout set with the signal-to-distortion ratio (SDR) as the figure of merit \cite{vincent2007first} (Table~\ref{tab:gamma}).

\begin{table}
    \centering
    \begin{tabular*}{\linewidth}{@{\extracolsep{\fill}}lccccccccc@{}}
        \toprule
         & learn & anechoic & 0 & 1 & 2 & 3 & 4 & 5 & 6 \\
         \cmidrule{2-10}
         $\gamma = $ & $10^{-1}$ & $10$ & $10$ & $10^{-3}$ & 0 & 0 & 0 & 0 & 0 \\
         \bottomrule
    \end{tabular*}
    \caption{Value of the regularization parameter $\gamma$ used with the universal dictionary.}
    \label{tab:gamma}
    \vspace{-0mm}
\end{table}

\begin{figure*}
    \centering
    \subfloat[mu_univ][MU-NMF, Universal dictionary]{\includegraphics[width=0.33\textwidth]{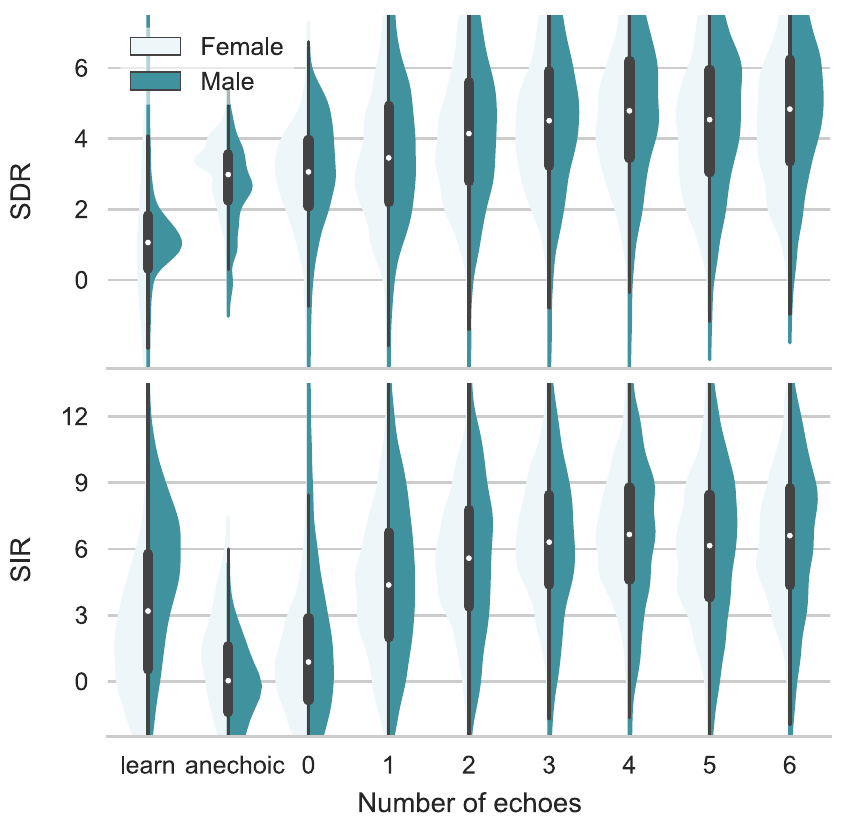}\label{fig:mu_univ}}
    \subfloat[mu_spkr][MU-NMF, Speaker-specific dictionary]{\includegraphics[width=0.33\textwidth]{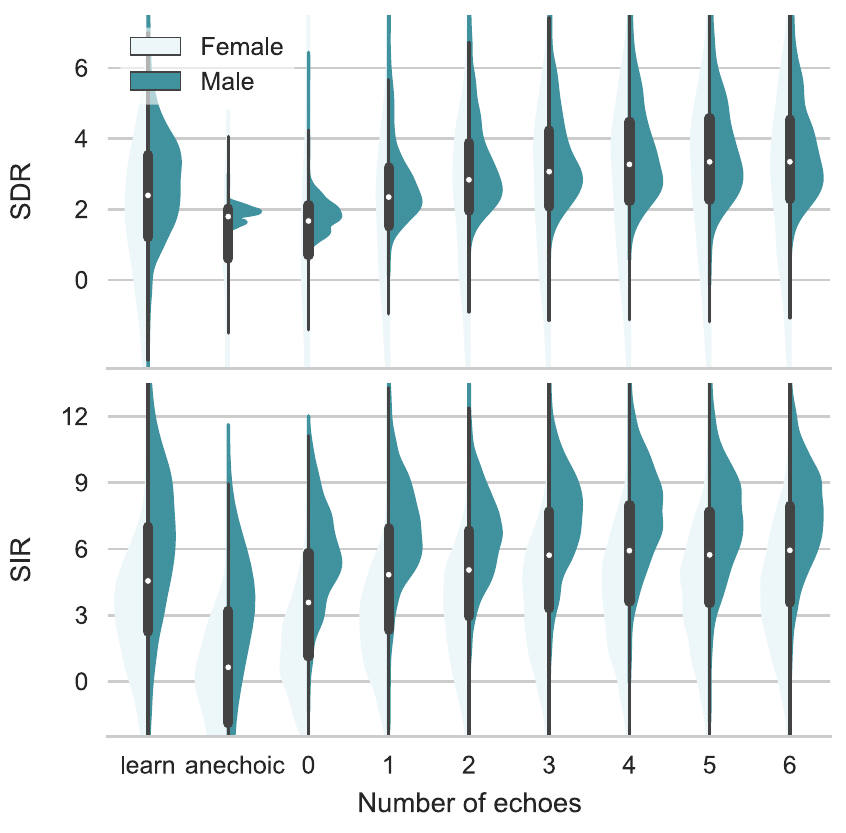}\label{fig:mu_spkr}}
    \subfloat[em_spkr][EM-NMF, Speaker-specific dictionary]{\includegraphics[width=0.33\textwidth]{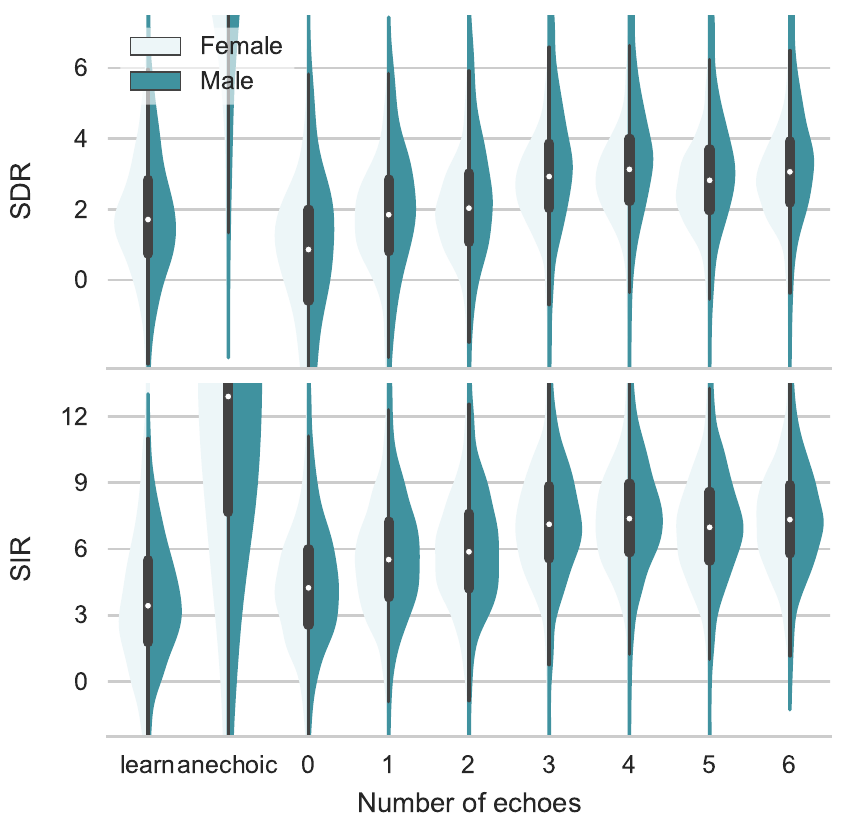}\label{fig:em_spkr}}
    \caption{Distribution of SDR and SIR for male and female speakers as a function of the number of echoes included in modeling, and comparison with the three baselines.}
    \label{fig:results}
    \vspace{-0mm}
\end{figure*}

\begin{figure}
    \centering
    \includegraphics{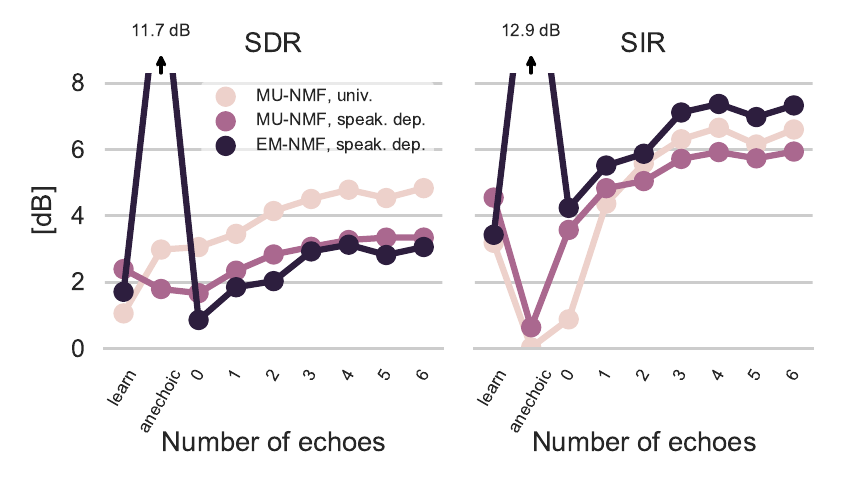}
    \caption{Summary of the median SDR and SIR for the different algorithms evaluated.\vspace{-3mm}}
    \label{fig:median}
\end{figure}

\subsection{Dictionary Training, Test Set, and Implementation}

\textit{Universal Dictionary:} Following the methodology of \cite{Sun:2013co} we select 25 male and 25 female speakers
and use all available training sentences to form the universal dictionary
$
    \mD = [\mD_1^\mathsf{M}\cdots \mD_{25}^\mathsf{M} \, \mD_{1}^\mathsf{F}\cdots\mD_{25}^\mathsf{F}].
$
The test signals were selected from speakers \emph{and} utterances outside the training set.
The number of latent variables per speaker is 10 so that with STFT frame size of 2048 we have $\mD\in\R^{1025\times500}$.

\textit{Speaker-Specific Dictionary:}  Two dictionaries were trained on one male and one female speaker. One utterance per speaker was excluded to be used for testing. The number of latent variables per speaker was set to 20.

All dictionaries were trained on samples from the TIMIT corpus \cite{garofolo1993timit} using the NMF solver in \textit{scikit-learn} Python package~\cite{pedregosa2011scikit}. 

\textit{Implementation:} Authors of \cite{ozerov2010multichannel} provide a Matlab implementation of MU-NMF and EM-NMF methods for stereo separation. We ported their code to Python and extended it to arbitrary number of input channels.\footnote{Our implementation and all experimental code are publicly available in line with the philosophy of reproducible research.} The number of iterations for MU-NMF (EM-NMF) was set to 200 (300) and simulated annealing in EM-NMF implementation was disabled.

\subsection{Results}
\label{sec:results}

We evaluate the performance in terms of signal-to-distortion ratio (SDR) and source-to-interference ratio (SIR) as 
defined in \cite{vincent2007first}. We compute these metrics using the \textit{mir\_eval} toolbox~\cite{raffel2014mir_eval}.

The distributions of SDR and SIR resulting from separation using MU-NMF and a universal dictionary are shown in Figure~\ref{fig:mu_univ}, with a summary in Figure~\ref{fig:median}. We use the median performance to compare the results
from different algorithms.
First, we confirm that separation fails for flat TFs (\textit{anechoic} and $K=0$) with SIR around 0~dB. Learning the TFs performs somewhat better in terms of SIR than in terms of SDR, though both are low. Introducing approximate TFs dramatically improves performance: we outperform the learned approach even with a single echo. With up to six echoes, gains are +2~dB SDR and +5~dB SIR. Interestingly, with more than one echo, $\ell_1$ regularization becomes unnecessary; non-negativity and echo priors are sufficient to separate sources.

Separation with speaker-dependent dictionaries is less challenging since we have a stronger prior. Accordingly, as shown in Figures~\ref{fig:mu_spkr} and \ref{fig:median}, MU-NMF now achieves some separation even without the channel information. The gains from using echoes are smaller, though one echo is still sufficient to match the median performance of learned TFs. Using an echo, however, results in a smaller variance. This is surprising at least to the authors of this paper. Adding more echoes further improves SDR (SIR) by up to +2~dB (+3~dB).

In the same scenario, EM-NMF (Figure~\ref{fig:em_spkr}) has near-perfect performance on anechoic signals which is expected as the problem is overdetermined. As for MU, a single echo suffice to reach the performance of learned TFs and more further improves it. Moreover, echoes
significantly improve separation power as
illustrated by up to 3~dB improvement over \textit{learn}. It is interesting to note that in all experiments the first three echoes almost saturate the metrics. This is good news since higher order echoes are hard to estimate.

\vfill
\section{Conclusion}

In this paper we began studying the role of early echoes in computational auditory scene analysis, in particular in source separation. We found a simple echo model not only improves performance, but it enables separation in conditions where it is not normally possible, for example with certain non-negative speaker-independent models. Echoes seem to play an essential role in magnitude-only algorithms like non-negative matrix factorization via multiplicative updates. They improve separation as measured in terms of standard metrics even when compared to approaches that learn the transfer functions. We believe these results are only a first step in understanding the potential of echoes in computational auditory scene analysis. They suggest that simple models used in this paper could be used as regularizers in other common audio processing tasks. Ongoing work includes running real experiments, studying the underdetermined case, and blindly estimating the wall parameters.



\clearpage

\bibliographystyle{IEEEtran}
\bibliography{refs}

\end{document}